\documentclass[longauth,letter,traditabstract]{aa} %\documentclass[referee]{aa}  
\usepackage{graphicx}   
\usepackage{natbib}   
\usepackage{color}   
\bibpunct{(}{)}{;}{a}{}{,} % to follow A&A style   
\newcommand{\micron}{\mbox{$\mu$m}}   

\newcommand{\HI}{\ion{H}{i}}   
\newcommand{\CI}{\ion{C}{i}}   
\newcommand{\CII}{\ion{C}{ii}}   
   
\newcommand{\msun}{M$_{\odot}$}   
   
\newcommand{\ratio} {N({\rm H}_2) / I_{\rm CO(1-0)}}
\newcommand{\ratioo} {N({\rm H}_2) / I_{\rm CO}}
\newcommand{\ratiot} {N({\rm H}_2) / I_{\rm CO(2-1)}}

\begin{document}   

\title{Cool gas and dust in M33: Results from the
{\it  Herschel} M33 extended survey (HERM33ES). \thanks{{\it Herschel} is an ESA space observatory with science instruments
provided by European-led Principal Investigator consortia and with
important participation from NASA.}}

  \author{  
    J.\,Braine\inst{1} \and
    P.\,Gratier\inst{1} \and
    C.\,Kramer\inst{2} \and
    E.M.\,Xilouris\inst{3}
    E.\,Rosolowsky\inst{4} \and
    C.\,Buchbender\inst{2}  \and
    M.\,Boquien\inst{5} \and
    D.\,Calzetti\inst{5} \and 
    G.\,Quintana-Lacaci\inst{2}  \and
    F.\,Tabatabaei\inst{6} \and
    S.\,Verley\inst{7} \and
    F.\,Israel\inst{8} \and 
    F.\,van der Tak\inst{9} \and
    S.\,Aalto\inst{10} \and
    F.\,Combes\inst{11} \and
    S.\,Garcia-Burillo\inst{12} \and
    M.\,Gonzalez\inst{2}  \and
    C.\,Henkel\inst{6} \and 
   B.\,Koribalski\inst{13} \and 
    B.\,Mookerjea\inst{14} \and
    M.\,Roellig\inst{15} \and
   K.F.\,Schuster\inst{16} \and
%    T.\,Wiklind\inst{17} \and
    M.\,Rela\~no\inst{18} \and
    F.\,Bertoldi\inst{19} \and
    P.\,van der Werf\inst{8} \and
    M.\,Wiedner\inst{11}
         }   

  \institute{       
  Laboratoire d'Astrophysique de Bordeaux, Universit\'{e} Bordeaux 1, 
    Observatoire de Bordeaux, OASU, UMR 5804, CNRS/INSU, B.P. 89, 
    Floirac F-33270
    \and %no2:
    Instituto Radioastronomia Milimetrica (IRAM), 
    Av. Divina Pastora 7, Nucleo Central, E-18012 Granada, Spain
    \and %no3:
    Institute of Astronomy and Astrophysics, National Observatory of Athens, 
    P. Penteli, 15236 Athens, Greece
    \and %no4:
    University of British Columbia, Okanagan, 3333 University Way, Kelowna BC V1V 1V7 Canada
    \and %no5
    Department of Astronomy, University of Massachusetts, Amherst, 
    MA 01003, USA 
    \and %no6
    Max-Planck-Institut f\"ur Radioastronomie (MPIfR), 
    Auf dem H\"ugel 69, D-53121 Bonn, Germany
%    \and %no3
    Argelander Institut fr Astronomie. Auf dem H\"ugel 71, 
    D-53121 Bonn, Germany
    \and %no7:
    Dept. F\'{i}sica Te\'{o}rica y del Cosmos, Universidad de Granada, Spain
    \and %no8:
   Leiden Observatory, Leiden University, PO Box 9513, NL 2300 RA Leiden, The Netherlands
    \and %no9:
    SRON Netherlands Institute for Space Research, Landleven 12, 
    9747 AD Groningen, The Netherlands
    \and %no10
    Onsala Space Observatory, Chalmers University of Technology, 
    43992 Onsala, Sweden
  \and %no11
   Observatoire de Paris, LERMA, 61 Av. de l'Observatoire, 
    75014 Paris, France
%    \and %no8:
%    European Southern Observatory, Casilla 19001, Santiago 19, Chile
    \and %no12:
    Observatorio Astron\'{o}mico Nacional (OAN) - Observatorio de Madrid, 
    Alfonso XII 3, 28014 Madrid, Spain
    \and %no13:
    Australia Telescope National Facility, CSIRO, PO Box 76, Epping, 
    NSW 1710, Australia
 %   \and %no13:
 %   Instituto de Estructura de la Materia (CSIC)
 %   Departamento de Astrofisica Molecular e Infrarroja
 %   C/ Serrano 123, E-28006 Madrid, SPAIN
    \and %no14:
    Department of Astronomy \& Astrophysics, 
    Tata Institute of Fundamental Research, 
    Homi Bhabha Road, Mumbai 400005, India
    \and %no15:
    KOSMA, I. Physikalisches Institut, Universit\"at zu K\"oln,   
    Z\"ulpicher Stra\ss{}e 77, D-50937 K\"oln, Germany    
    \and %no16:
    IRAM, 300 rue de la Piscine, 38406 St. Martin d'H\`{e}res, France
%    \and %no17:
%    Space Telescope Science Institute, Baltimore MD 21218, USA
    \and %no18:
    Institute of Astronomy, University of Cambridge, Madingley Road, 
    Cambridge CB3 0HA, England
    \and %no19:
    Argelander Institut fr Astronomie. Auf dem H\"ugel 71, 
    D-53121 Bonn, Germany
%   \and %no19:
%    California Institute of Technology, MC 105-24, 1200 East 
%    California Boulevard, Pasadena, CA 91125, USA
%    \and %no20:
%    Department of Astronomy, Cornell University, Ithaca, NY 14853, USA
%    \and %no22:
%    Joint Astronomy Centre, 660 North A'ohoku Place, University Park, 
%    Hilo, HI 96720, USA
%   \and %no25:
%    Observatoire de Paris, 61 Ave. de l'Observatoire, F-75014 Paris, France
}   

  \offprints{J.\,Braine, \email{Braine@obs.u-bordeaux1.fr}}   
  \date{Received / Accepted 11 May, 2010}   

% context, aims, methods, results, conclusions  
  \abstract{We present an analysis of the first space-based far-IR-submm observations of M~33, which measure the emission from the cool dust and resolve the giant molecular cloud complexes.  With roughly half-solar abundances, M33 is a first step towards young low-metallicity galaxies where the submm may be able to provide an alternative to CO mapping to measure their H$_2$ content.  In this Letter, we measure the dust emission cross-section $\sigma$ using SPIRE and recent CO and \HI\ observations; a variation in $\sigma$ is present from a near-solar neighborhood cross-section to about half-solar with the maximum being south of the nucleus.  Calculating the total H column density from the measured dust temperature and cross-section, and then subtracting the \HI\ column, yields a morphology similar to that observed in CO.  The H$_2$/\HI\  mass ratio decreases from about unity to well below 10\% and is about 15\% averaged over the optical disk.  The single most important observation to reduce the potentially large systematic errors is to complete the CO mapping of M~33. }
  \authorrunning{Braine et  al.} 
%  \titlerunning{}   
  \maketitle   
%________________________________________________________________   

 % \section{Introduction } %%%%%%%%%%%%%%%%%%%%%%%%%%%%%%%%%%%%%%%%%%%
%%

%  Within the framework of the open time key project ``Herschel M33
%  extended survey ({\tt HERM33ES})'', we are studying the unique,
%  nearby, metal poor, face-on galaxy M33 to understand the origin of
%  various diagnostic lines and heating/cooling and other processes in
%  the interstellar medium (ISM).

\begin{figure}[h!]   
 \centering   
 \includegraphics[width=8.8cm,angle=0]{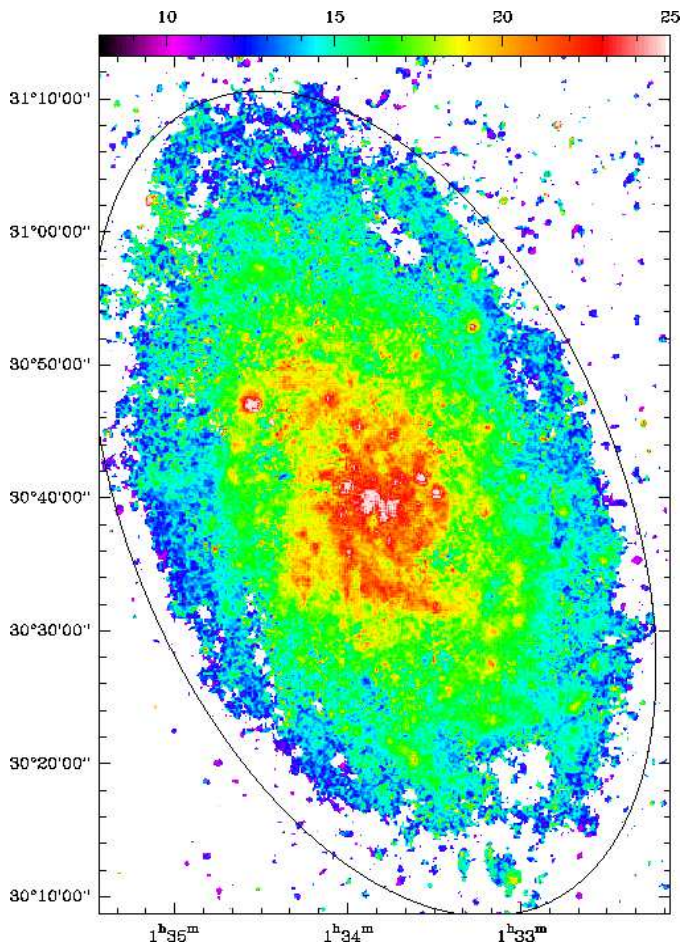} 
 \caption{Dust color temperature map of M~33 calculated from the SPIRE 250 and 350 \micron\ flux ratio with $\beta=2$ at the $\sim 25''$ resolution of the 350\micron\ data.
 Coordinates are J2000, the temperature scale is shown at the top in Kelvin, and the ellipse shows the 8 kpc radius, slightly beyond R$_{25}$. }
\label{fig-maps}   
\end{figure}   

\section{Introduction, data, and dust temperature}

Understanding star formation requires studying the interplay between the phases of the interstellar 
medium (ISM).  Dust processes most of the energy transiting the ISM, but the cool dust component, although representing the vast majority of the dust mass, is difficult to observe from the ground.
{\it Herschel} SPIRE observations \citep{Pilbratt10,Griffin10} are the first space-based 250-500 \micron\ data and as 
such provide a unique occasion to put together a global picture of the cool gas and dust in M~33.  In particular, we 
compare the morphology of the Far-IR emission and that of the gas as determined from CO and \HI\ measurements
and attempt to measure how the dust cross-section varies in M~33.
A longer term goal is to be able to use the dust emission to constrain the variation of the $\ratioo$ factor within M~33
and elsewhere.

This Letter is one of a series on the {\tt HERM33ES} project on the ISM of the Local Group galaxy M~33, an overview of which is given in \citet{Kramer10}, hereafter K10. For consistency with the other M\,33 papers in this volume, we 
  adopt a distance of $D$ = 840 kpc for M\,33 (i.e., 25 arcsec = 100 pc) 
  and orientation parameters of $PA$ = 22.5 degrees and $i$ = 56\degr. We use the recent  {\tt HERM33ES} SPIRE observations at 250, 350, and 500 \micron\, combined with CO(2--1) observations from the M33CO@IRAM project, as a tracer of the 
molecular component, and a high-resolution mosaic of VLA HI data \citep[both from][]{Gratier10}.  
The SPIRE data were first processed as described in K10 and then
converted from Jy/beam to brightness units (MJy/sr).  
To estimate the dust temperature, the 250\micron\ data were convolved to the 350\micron\ beamsize ($\sim 25"$) and, assuming a single temperature grey body with an emissivity $\propto \nu^\beta$ with $\beta =2$, a temperature was derived from the flux ratio.
At the temperatures of the cool component seen in M~33, \citet{Dupac03} find $\beta \approx 2$.
To minimize the effect of the uncertainty in $\beta$, we chose to use adjacent bands to estimate temperature. 
The 250/350\micron\ ratio provides more accurate temperatures than the 350/500 
micron ratio even for temperatures below 10~K.  A 15\% variation (or uncertainty) in the 250/350\micron\
ratios corresponds to a temperature change of 0.9, 2.1, and 4.1 K at temperatures of 10, 15, and 20 K
but the same uncertainty in the 350/500\micron\ ratios yields 1.3, 3.2, and 6.3 K errors for the same 
dust temperatures.  A further advantage is that we obtain the dust temperature at a resolution typical of giant molecular clouds (GMC), $\sim 100$ pc.

Ideally, a multi-component fit would be used but this requires high S/N data at many wavelengths over the whole disk.  By assuming that the cool dust component dominates the
emission beyond 250\micron, we could calculate the dust temperature out to the optical radius of M~33.  The resulting temperature map is shown in Fig. 1.  Figure 2 shows a comparison with the temperature of the cool component of the preferred two-component (warm plus cool dust) model fit to data between 24\micron\ and 500\micron\ from K10.  
The two-component model uses $\beta=1.5$ and the temperatures are higher out to 6~kpc.  
%However, the temperature of the cool component depends not only on the value of $\beta$ used but also on the temperature and mass of the "warm" component(s).
%In the inner 3.5 kpc, the two-component model yields dust temperatures up to 2 K cooler and thus masses about 15\% higher.  We attribute the higher temperatures beyond 4 kpc to the incomplete coverage of the two-component fit such that only the brighter, usually warmer, regions are included in the average.
The temperatures in the radial bins have been estimated by averaging the temperatures rather than averaging the emission as in K10.  The latter yields slightly higher temperatures because the dust is usually warmer where emission is strong (compare Fig. 1 with Fig 1. in K10).  
K10 calculate the dust column density using $\kappa=0.4 (\nu/250{\rm GHz})^2 \ {\rm cm}^2$g$^{-1}$ of 
dust, which is equivalent to a dust cross-section per H-atom of $\sigma = 1.1 \times 10^{-25} (\lambda/250\mu {\rm m})^{-2} 
{\rm cm}^2$ for a hydrogen gas-to-dust mass ratio of 140.  
In the following we use the SPIRE 250/350 \micron\ color temperature because the data cover a greater area and agree well with the cool dust temperature of the two-component fit, showing the domination of the cool dust at these wavelengths.
No correction for line contamination was subtracted from the SPIRE data.  At these frequencies, the CO lines
contribute very little to the continuum flux, unlike at 1.3mm or 850\micron\ \citep[e.g.][]{Braine_n3079}.  

Figure 3 shows the SPIRE 250\micron\ emission with the CO(2--1) emission as contours at 25$''$ resolution to show how closely the CO emission follows the dust emission peaks. \citet{Gratier10} show a detailed comparison of the CO and \HI\ emission with star formation tracers such as H$\alpha$ and {\it Spitzer} 8 and 24\micron\ maps.
 The  250\micron -bright regions are detected in CO and the general morphology of the cool dust emission is seen in the \HI\ image.  

\begin{figure}[h]   
 \centering   
 \includegraphics[width=6.2cm,angle=-90]{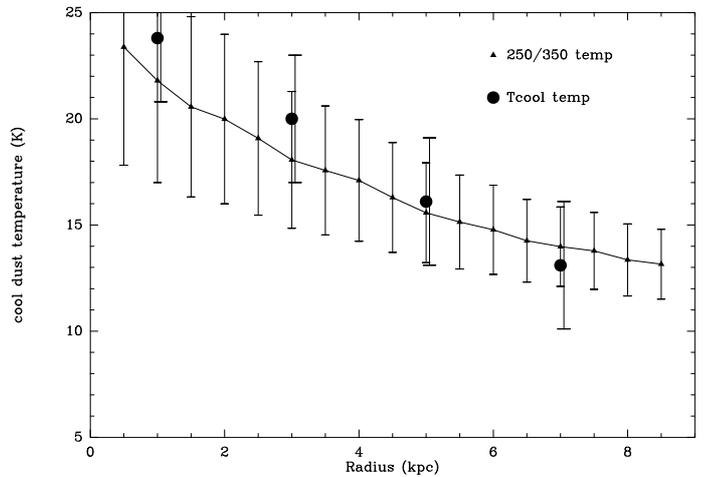} \\ 
 \caption{Radial distribution of the average dust temperature in 
  the stellar disk of M\,33. The solid line joining the triangles shows the 250/350 \micron\ color temperature
and for comparison we show the cool dust temperature (large dots) 
  obtained by K10 in their preferred two-component model with $\beta=1.5$.
The error bars on the triangles 
  indicate a 15\% calibration uncertainty in the 250/350 \micron\ ratio converted to temperature.}
% The two fits yield quite consistent results.
\end{figure}   

\begin{figure}[h]   
 \centering   
 \includegraphics[width=8.8cm,angle=0]{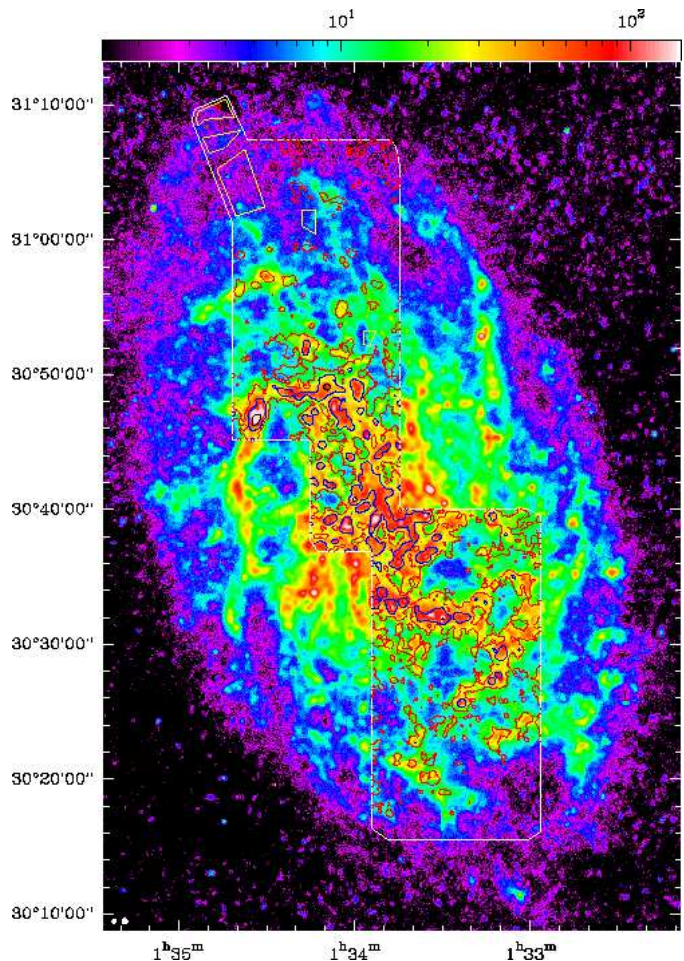} \\ 
 \caption{SPIRE 250\micron\ emission with the IRAM CO(2--1) emission at 25$''$ resolution superposed as contours and with the polygons to the upper left used to calculate the dust cross-section per H-atom.
 The  beams are shown as white dots in the lower left corner.  The CO contours are at 0.5 (red), 2 (blue), 
 and 4 K km s$^{-1}$ in the main beam scale.  The large polygon indicates the area covered so far by the IRAM CO survey.}
\end{figure}   

%\begin{figure}[h]   
% \centering   
% \includegraphics[width=8.8cm,angle=0]{s500_co_fig.eps} \\ 
% \caption{SPIRE 500\micron\ emission with the IRAM CO(2--1) emission at 25$''$ resolution superposed.  
% The beams are shown as white dots in the lower left corner.  The CO contours are at 0.4 (red), 0.8 (blue), 
% 1.6, and 3.2 K km s$^{-1}$ in the main beam scale. The polygon indicates the area covered so far by the IRAM CO survey.}
%\end{figure}   

\section{Dust cross-section}

In order to estimate the total gas mass of a galaxy from its dust emission, it is first necessary to measure the dust emission cross-section 
per H-atom $\sigma$.  We do this by selecting regions with HI emission and well-constrained
dust temperatures but with little or no CO emission (in order to avoid the assumption of a $\ratioo$ value and CO line ratio).  Thereby, we can relate the observed neutral hydrogen column density, N$_{\rm HI}$, to the dust optical depth, $\tau = \sigma {\rm N_H} = S_\nu /B_{\nu,T}$, and infer the cross section $\sigma$.  In practice, two methods were used: ($i$) we selected large contiguous regions as shown on the map in Fig. 3 and ($ii$) all pixels with HI emission, a defined dust temperature, and no detected CO emission (I$_{\rm co} < 0.05$ K km s$^{-1}$) were taken, using maps at the resolution of the 350\micron\ SPIRE data.  
%Since the dust temperature is defined over a large region (Fig. 1), this does not reduce the area that can be used. 
In the optically thin limit,
\begin{equation}
$$ N_{\rm tot} = N_{\rm HI} + N_{\rm H_2} = S_\nu / (\sigma  \, \, B_{\nu,T}) $$
\end{equation}
where $S_\nu$ is the flux density and $N_{\rm tot}$ is the H column density.  Where CO is not detected (i.e. 
$N_{\rm tot} \approx N_{\rm HI}$), we can calculate $\sigma$ from the dust temperature and the observed \HI\ column density and 
SPIRE flux by $\sigma \approx S_\nu / (N_{\rm HI}  \, \, B_{\nu,T}) $.  
%The optically thin assumption is shown to be justified in \citet{Xilouris10}.

From the regions indicated in Fig. 3, we estimate $\sigma$  (using Eq. 1) for different radii and the results are shown in the first line of Table 1.  We assume that the cross section
within the inner 4 kpc is the same as at 4kpc for two reasons: ($i$) there are few truly CO-free regions in the central part of M~33 and ($ii$) \citet{Gardan07} showed (their Fig. 13) that the \HI-H$_2$ relation was similar within 4kpc but beyond 4 kpc the H$_2$ fraction decreased sharply even at constant total Hydrogen column density, so we consider $R\sim 4$kpc as a transition in ISM properties.  It is very likely that some molecular gas is present in these regions, particularly those closer to the center, so $\sigma$ is likely overestimated.  In 
the second approach, in order to exclude averaging in undetected molecular clouds, we take the peak of the histogram of $\sigma$ values for the areas without detected CO emission but with HI emission and a constrained dust temperature, yielding the lower values for $\sigma$, as shown in Table 1.
Using this method, we find a north-south difference in M33 at equivalent radii.  
The uncertainty in the histogram method is about $\pm 0.1$; although the distribution is sometimes broad, 
the peak value is well defined.  The histograms of $\sigma$ values are clearly different in the north and south.
The polygon method was difficult to use in the south due to the absence of large CO-free regions with HI and dust emission.  An advantage of the histogram method is that it excludes the tail of high $\sigma$ which may be due to H$_2$ without detected CO emission.

The intrinsic expectation is that because the oxygen abundance is about half solar, the dust cross-section 
should be as well \citep{Draine07}, and that given 
the shallow abundance gradient in M33 \citep{Rosolowsky07,Magrini09}, this should hold for the entire galaxy.  However, it quickly became
apparent that the dust cross-section per H-atom varied over the galaxy, decreasing with radius by close to a factor 2, with a higher abundance in the south -- similar to the variation found by \citet{Magrini10} from optical \ion{H}{II} region and PNe abundance measurements.  In the solar neighborhood, $\sigma$ is about $\sigma \approx 1.1 \times 10^{-25}$cm$^2$
per H-atom \citep{dl84,Draineli07} at 250$\mu$m and varying as $\nu^2$.  If we apply a $\sigma$ of half this value to M33 using the dust temperature map in Fig. 1, we obtain unrealistically high gas masses in the inner disk: $\sim 40$ \msun\ pc$^{-2}$ on average, whereas the \HI\ is about 11 \msun\ pc$^{-2}$ and the H$_2$ considerably less \citep{Gratier10}. 
For a half-solar $\sigma$, the (cool) dust temperature would have to be about 30K, beyond any experimental uncertainties, so we conclude that $\sigma$ must be higher in the inner disk. 
 In the outer disk, however, the \HI\ column density (where no CO emission is present) is roughly equal to what is derived from the dust with $\sigma = 0.5 \sigma_\odot$, suggesting that this value is appropriate for the outer disk.  

\begin{table}[h*]   
 \caption[]{ Dust cross-section $\sigma$ at 250\micron\ as a function of radius in M33, expressed in units of $1.1 \times 10^{-25}$cm$^2$ per H-atom.  The polygons are shown in Fig. 3.  The histogram method was applied separately in the north and south (lines 2 and 3), leading to the "Model" values used to estimate the total H column density to make Fig. 4.}   
\begin{tabular}{lrrrrrr}   
\hline \hline  
r (kpc) & 4 & 5& 5.5 &6& 7& 7.5 \\
\hline
Polygons & 1.8 & 1.02 &..& 1.07 & 0.66 & 0.50  \\
histo-N   & 0.65 &..& 0.54 & ..& ..&0.48 \\
histo-S   & 0.92 &..& 0.95 & 0.69&.. &.. \\
% radius & $<5$ & $<6$ & $<7$ & $<7.5$ \\
Model   & 0.8 &..& 0.75 &..& 0.66 & 0.5 \\
\end{tabular}   
\end{table}

%\begin{figure}[h!]   
% \centering   
% \includegraphics[width=8.8cm,angle=-0]{nhi_poly_co_fig.eps} 
% \caption{Grey-scale map of \HI\ column density with the polygons to the upper left used to calculate the dust cross-%section per H-atom.  The large polygon is the area covered by the CO(2--1) map with
% CO(2--1) contours at 25$"$ resolution superposed \citep{Gratier10}.
% The ellipse shows the 8 kpc radius, slightly beyond R$_{25}$, for a distance of 840 kpc, a position angle of 22.5$^\circ$, and an inclination of 56$^\circ$. }
%\label{fig-maps}   
%\end{figure}   

\begin{figure}[h!]   
 \centering   
 \includegraphics[width=8.8cm,angle=0]{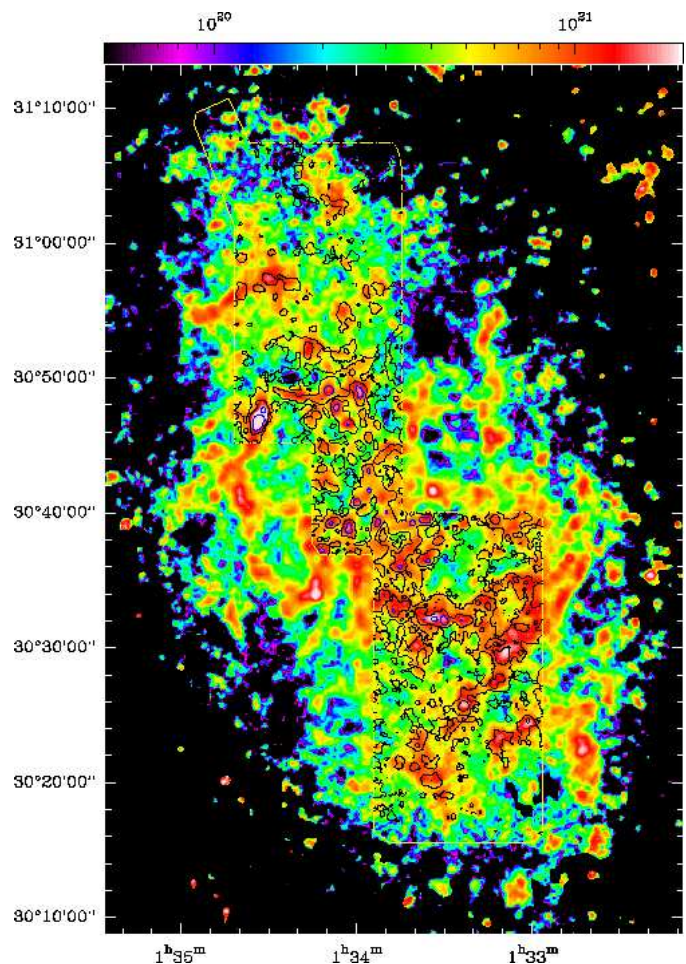} 
 \caption{H$_2$ column density map estimated from the 500\micron\ SPIRE map, $\beta=2$, the 250/350\micron\ dust temperature as shown in Fig. 1, the \HI\ data from \citet{Gratier10}, and the cross-sections given in Table 1.  The column density scale is shown at the top in units of H$_2$ molecules per cm$^2$.  CO contours at 0.4 and 1.6 K km/s (black) and 5 K km/s (blue) are superposed. }
\label{fig-maps}   
\end{figure}   

\begin{figure}[h!]   
 \centering   
 \includegraphics[width=9cm,angle=0]{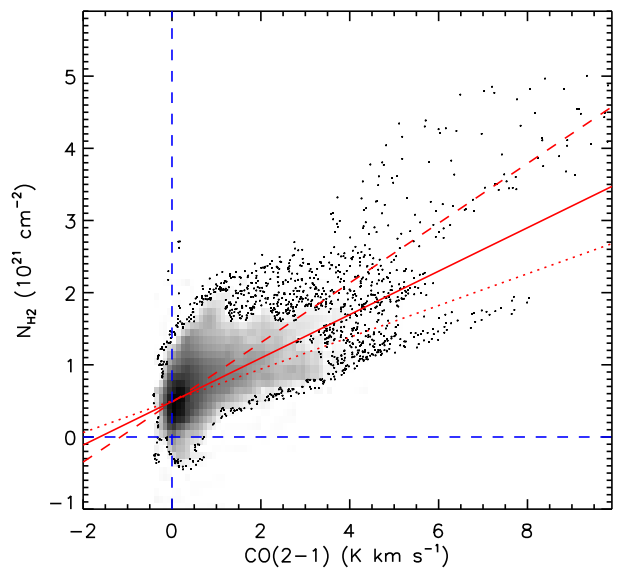} 
 \caption{Scatter plot of H$_2$ column density as derived from the dust emission and the CO(2-1) intensity.
 The solid red line is a fit to the entire dataset and corresponds to a $\ratiot$ value of $3 \times 10^{20}$ H$_2$ per K km s$^{-1}$ with an offset as described in the text.  A radial variation is present: fitting within 2 kpc yields a $\ratiot$ value of $2.2 \times 10^{20}$ H$_2$ per K km s$^{-1}$  and $4.1 \times 10^{20}$ H$_2$ per K km s$^{-1}$ beyond 2 kpc (dotted and dashed lines, respectively).} 
\label{fig-maps}   
\end{figure}

\section{Total and molecular gas mass}

To really obtain an accurate gas mass from the cold dust emission data, the calibration uncertainties in the SPIRE bands must be reduced because the 15\% uncertainty is not sufficient to distinguish between different solutions of $\beta$ and temperature.  A second requirement is that the whole disk of M33 must be observed in CO so that a well-determined variation of $\sigma$, based on the second technique, can be constructed.  The \HI\ data provide a reliable picture of the \HI\ column density, so subtracting this from the total gas column density derived from the dust emission yields an estimate of the molecular gas mass which does not rely on CO emission.  A further caveat is that $\sigma$ must not change from the atomic to molecular medium.  Once the [\CI ] and [\CII ] lines have been measured, these uncertainties can be reduced by the additional constraint that the C/H ratio (from the sum of the C reservoirs)
should be equal to the C abundance derived by other means.

Using the information currently available, we have used Table 1 (second method) to build an image of the dust cross-section $\sigma$, erring on the low-$\sigma$ side, as indicated in the last line of Table 1.   
We then used this $\sigma$, scaled to 500\micron\ with $\beta=2$, and the dust temperature (Fig. 1) to estimate the total gas mass from the 500\micron\ data and then subtracted the \HI\ column density to obtain the H$_2$ column density using the interferometric \HI\ maps from \citet{Gratier10}, which recover more than 90\% of the flux found by \citet{Putman09} using Arecibo.  The missing 21cm flux is expected to be located in the mid-to-outer disk where the rotation curve is fairly flat.  A morphological comparison of the dust-derived H$_2$ column and the CO contours in Fig. 4 shows that while major regions were not missed by the CO observations, the dust-derived H$_2$ map 
does not correlate perfectly with the CO.
%shows the same differences with respect to the CO map as did the FIR observations themselves \citep[see][]{Gratier10}.  
For example, the strong CO peak at (01$^h$34$^m$09$^s$.4,$+$30$^\circ$49$'$06$''$) corresponds to a red (not white) region in Fig. 4; this may of course reflect a real variation in the $\ratioo$ ratio.

The data used in the scatter plot in Fig. 5 \citep[see also Fig. 7 in ][]{Leroy09} yield a formal fit with a 
$\ratiot=3 \times 10^{20}$ H$_2$ per K km s$^{-1}$ value, corresponding to a roughly galactic 
$\ratio = 2.1 \times 10^{20}$ H$_2$ per K km s$^{-1}$ for a line ratio CO(${{2-1} \over {1-0}}$) $ = 0.7$.  
The inner disk $\ratiot$ is lower and rises to about $4 \times 10^{20}$ H$_2$ per K km s$^{-1}$
beyond 2 kpc.  The offset, in principle, implies that some low column density H$_2$ is not seen in CO.
Both values depend strongly on the details of the dust cross-section, illustrating the need for a complete map of M 33 in CO.  
%because a small difference in $\sigma$ yields a an equivalent change in total H column 

The total hydrogen mass derived from the dust emission within 8kpc is estimated to be $1.63 \times 10^9$ M$_\odot$, very similar to the more directly measured \HI$+$H$_2$ masses in \citet{Gratier10}. 
%If we use the cool dust temperatures from the two-component fit for the inner 3.5 kpc, the estimated Hydrogen mass increases by $\sim 7 \times 10^7$ M$_\odot$, to $1.7 \times 10^9$ M$_\odot$.  
The dust cross-sections we derive suggest dust-to-gas mass ratios (including He) ranging from $\sim 125$ in the inner 4~kpc to 200 near R$_{25}$ for Milky Way like dust ({\it cf.} K10).  Comparing the total dust-derived H$_2$ mass with the CO 
luminosity \citep{Gratier10} yields a $\ratioo$ of about 1.5 times the Galactic $\ratioo$.  
Because the \HI\ dominates, however, a small error in the total dust-derived gas mass translates into a large uncertainty in the dust-derived H$_2$ mass.  It is clearly necessary to obtain whole-galaxy measurements of $\sigma$ in CO-free zones to refine these estimates.

%\begin{acknowledgements} 
%%%%%%%%%%%%%%%%%%%%%%%%%%%%%%%%%%%%%%%%%%%%%%%%  

%\end{acknowledgements}   

\bibliographystyle{aa} %%%%%%%%%%%%%%%%%%%%%%%%%%%%%%%%%%%%%%%%%%%%%%%%%%   
\bibliography{../jb} % herm33es-01.bib

\end{document}